
\input harvmac.tex
%
%
\def\nopage0{\pageno=0 \footline={\ifnum\pageno<1 {\hfil} \else
              {\hss\tenrm\folio\hss}   \fi}}
%
%
\def\pagenumbers{\footline={\hss\tenrm\folio\hss}}
%
%
\def\sqr#1#2{{\vcenter{\hrule height.#2pt
      \hbox{\vrule width.#2pt height#1pt \kern#1pt
         \vrule width.#2pt}
      \hrule height.#2pt}}}

%
%
\def \d2dots{\mathinner{\mkern1mu\raise1pt\vbox{\kern7pt\hbox{.}}\mkern2mu
\raise4pt\hbox{.}\mkern2mu\raise7pt\hbox{.}\mkern1mu}}
%
%
\def\smdrct#1{{\vcenter{\hbox{\vrule width.4pt height#1pt}}\kern-1.5pt\times}}

%
%

%
%

                    \def\cR{{\cal R}}
                    \def\cU{{\cal U}}

%
%

\def\CC{\rlap {\raise 0.4ex \hbox{$\scriptscriptstyle |$}}
\hskip -0.1em C}
\def\FF{\hbox to 8.33887pt{\rm I\hskip-1.8pt F}}
\def\NN{\hbox to 9.3111pt{\rm I\hskip-1.8pt N}}
\def\PP{\hbox to 8.61664pt{\rm I\hskip-1.8pt P}}
\def\QQ{\rlap {\raise 0.4ex \hbox{$\scriptscriptstyle |$}}
{\hskip -0.1em Q}}
\def\RR{\hbox to 9.1722pt{\rm I\hskip-1.8pt R}}
\def\ZZ{\hbox to 8.2222pt{\rm Z\hskip-4pt \rm Z}} 
\def\demi{{1\over 2}}
\lref \rDAFus { D. Arnaudon, {\sl Fusion rules and $\cR$-matrix
for the composition of regular spins with semi-periodic representations of
$SL(2)_q$,} Phys. Lett. {\bf B268} (1991) 217.}
\lref \rDAslnfusion {D. Arnaudon, {\sl New fusion rules and
${\cal R}$-matrices for $SL(N)_q$ at roots of unity, }
Preprint CERN-TH.6324/91, to be published in Physics Letters {\bf B}.}
\lref \rAGS { L. Alvarez-Gaum\'e, C. G\'omez and G. Sierra,
{\sl Hidden quantum symmetries in rational conformal theories,} Nucl.
Phys. {\bf B319} (1989) 155;
{\sl Quantum group interpretation of some conformal field theories,} Phys.
Lett. {\bf B220} (1989) 142;  {\sl Duality and quantum groups,} Nucl. Phys.
{\bf B330} (1990) 347.}
\lref \rBS { V.V. Bazhanov and Yu.G. Stroganov,
{\sl Chiral Potts model as a
descendent of the six-vertex model,} J. Stat. Phys. {\bf 51} (1990) 799.}
\lref \rDK {  C. De Concini and V.G. Kac,
{\sl Representations of quantum groups at roots of 1,}
Progress in Math. {\bf 92} (1990) 471 (Birkh\"auser).}
\lref \rDJMM { E. Date, M. Jimbo, K. Miki and
T. Miwa,
{\sl New $\cR$-matrices associated with cyclic representations of
$\ \cU_q(A_2^{(2)})$}, preprint RIMS-706 and
{\sl Generalized chiral Potts models and minimal cyclic
representations of
$\ \cU_q \left(\widehat{gl}(n,\CC )\right) $,}
Commun. Math. Phys. {\bf 137}  (1991) 133.}
\lref \rGRS { C. G\'omez, M. Ruiz-Altaba and G. Sierra,
{\sl New $\cR$-matrix associated with finite dimensional representations
of $\ \cU_{q}(SL(2))$ at roots of unity,} Phys. Lett. {\bf B265} (1991) 95.}
\lref \rGS { C. G\'omez and G. Sierra,
{\sl A new solution  of the Star-Triangle equation based on $\cU _q(sl(2))$
at roots of unit,} Preprint CERN-TH.6200/91 and Gen\`eve  UGVA-DPT
1991/08-739.}
\lref \rKel { G. Keller,
{\sl Fusion rules of $\ \cU_{q}(SL(2,\CC))$, $q^{m}=1$,} Letters in
Math. Phys. {\bf 21} (1991) 273.}
\lref \rKMS { R.M. Kashaev, V.V. Mangazeev and  Yu.G.
Stroganov,  {\sl Cyclic eight-state $\cR$-matrix related to
$\cU_q(sl(3))$ algebra at $q^{2}=-1$}, Preprint IHEP (1991), and
{\sl $N^3$-state $\cR$-matrix related with $\cU _q(sl(3))$ algebra at
$q^{2N}=1$}, preprint RIMS-823 (1991).}
\lref \rMR { G. Moore and N. Reshetikhin, {\sl A comment on
quantum group symmetry in conformal field theory,} Nucl. Phys.  {\bf B328}
(1989) 557.}
\lref \rPas { V. Pasquier, {\sl Etiology of IRF models,}
Commun.
Math. Phys. {\bf 118}  (1988) 355.}
\lref \rPS { V. Pasquier and  H. Saleur, {\sl Common
structure between finite systems and conformal field theories through
quantum groups,} Nucl. Phys. {\bf B330}  (1990) 523.}
\lref \rRA {  P. Roche and  D. Arnaudon,
{\sl Irreducible representations of the quantum
analogue of $SU(2)$.}
Lett. Math. Phys. {\bf 17} (1989) 295.}
\lref \rRosA { M. Rosso,
{\sl Finite dimensional representations of the quantum analogue of the
enveloping algebra of a complex simple Lie algebra,}
Commun. Math. Phys. {\bf 117}, 581 (1988).}
\lref \rRuiz { M. Ruiz-Altaba, {\sl New solutions to the Yang--Baxter
equation from
two-dimensional representations of $\ \cU_q(sl(2))$ at roots of unit,}
Preprint UGVA-DPT 1991/08-741.}
\lref \rSkl { E. K. Sklyanin, {\sl Some algebraic
structures connected with the
Yang--Baxter equation. Representations of quantum algebras,}
Funct. Anal. Appl. {\bf 17} (1983) 273. }
\line{\hfil CERN-TH.6416/92}
\Title{}{\vbox{\centerline{Fusion Rules and $\cR$-Matrices}
      \vskip2pt\centerline{For Representations of $SL(2)_q$ at Roots of
       Unity*}} }
\footnote{}{*Talk given at the V$^{\rm th}$ Reg. Conf. on Math. Phys.,
Trakya University, Edirne, Turkey (1991) }

\centerline{Daniel Arnaudon\footnote{$^\dagger$}
{arnaudon@cernvm.bitnet}}
\footnote{}{On leave from Ecole Polytechnique, 91128 Palaiseau, FRANCE}

\bigskip
\centerline{Theory Division}
\centerline{CERN}
\centerline{1211 Gen\`eve 23, Switzerland}
\def\draftmode{\message{ DRAFTMODE }\def\draftdate{{\rm Version provisoire:
\number\day/\number\month/\number\yearltd\ \ \hourmin}}%
\headline={\hfil\draftdate}\writelabels\baselineskip=20pt plus 2pt minus 2pt
 {\count255=\time\divide\count255 by 60 \xdef\hourmin{\number\count255}
  \multiply\count255 by-60\advance\count255 by\time
  \xdef\hourmin{\hourmin:\ifnum\count255<10 0\fi\the\count255}}}
\def\Spin{{\rm Spin}\,}
\def\Ind{{\rm Ind}\,}
\def\TBi{{\rm B}\,}
\def\TBii{{\rm B'}\,}
\vskip .3in
We recall the classification of the
irreducible representations of $SL(2)_q$,
and then give
fusion rules for these representations. We also
consider the problem of $\cR$-matrices, intertwiners of the differently
ordered tensor products of these representations, and satisfying altogether
Yang--Baxter equations.
\vfill
\leftline{CERN-TH.6416/92}
\leftline{February 1992}
\leftline{hepth@xxx/9203011}
\tenpoint\supereject\global\hsize=\hsbody
\pagenumbers

\newsec{Introduction}
Representations of $SL(2)_q$ at roots of unity \refs{\rSkl, \rRA}
play several roles in physics.
The irreducible representations corresponding to deformations of classical
ones are used in conformal theory \refs{\rAGS, \rMR} and in
statistical theory \refs{\rPas, \rPS},  whereas the new periodic
representations (which exist only when $q$ is a root of unity) appear in
relation with statistical models
\refs{\rSkl, \rBS \rDJMM \rKMS \rGRS \rGS {--} \rRuiz}.
\medskip
We present here fusion rules for both types of representations, and
$\cR$-matrices that intertwine $\Delta$ and $\Delta'$ on  tensor products.
\medskip
The end of the introduction is devoted to definitions and we will also
recall the classification of the irreducible representations (irreps) of
$SL(2)_q$. In section 2, we recall the fusion rules of $q$-deformed irreps.
In sections 3 and 4, we consider the fusion rules involving the
representations that exist only when $q$ is a root of 1. Section 5 is
devoted to $\cR$-matrices.

\medskip
$SL(2)_{q}$ is defined by the generators $k$, $k^{-1}$, $e$, $f$,
and the
relations
\eqn\eSL{
\eqalign{
& kk^{-1}=k^{-1}k=1 ,\cr
& kek^{-1}=q^2 e ,\cr
& kfk^{-1}=q^{-2}f ,\cr
& [e,f]={k-k^{-1} \over q-q^{-1}}. \cr }}

The
coproduct $\Delta $ is given by
\eqn\eDelta {
\eqalign {
& \Delta(k)=k\otimes k \cr
& \Delta(e)=e\otimes 1      + k\otimes e \cr
& \Delta(f)=f\otimes k^{-1} + 1\otimes f \;,\cr  }}
while the opposite coproduct $\Delta '$ is $\Delta '= P \Delta P$ where $P$
is the permutation map $P x\otimes y =y \otimes x$.

The result of the composition of two representations
$\rho_{1}$ and $\rho_{2}$ of $SL(2)_{q}$ is the representation
$\rho=(\rho_{1}\otimes \rho_{2})\circ \Delta$, whereas the composition in
the reverse order is equivalent to
$\rho '=(\rho_{1}\otimes \rho_{2})\circ\Delta'$.

When $q$ is not a root of unity, the representation theory is similar to the
classical one \rRosA.

In the following, the parameter $q$ will be a root
of unity.  Let $m'$ be the smallest integer such that $q^{m'}=1$.
Let $m$ be
equal to $m'$ if $m'$ is odd, and to $m'/2$ otherwise.

In addition to the usual quadratic Casimir
\eqn\eCas{C=fe+(q-q^{-1})^{-2}\left( qk+q ^{-1}k^{-1} \right)}
the centre of $SL(2)_q$ contains also $e^m$, $f^m$, and $k^{\pm m}$.
Following ref. \rDK, we will denote by $x$, $y$, $z^{\pm 1}$, and $c$ the
values of $e^m$, $f^m$, $k^{\pm m}$, and $C$ on irreducible representations.

We now recall the classification \rRA\ of the irreducible representations of
$SL(2)_q$.
The new facts are that the
dimension of the finite dimensional irreps
are  bounded by $m$, and that the
irreps of dimension $m$ depend on three complex continuous parameters.
In the following, we will call type A irreps those that have a classical
analogue and type B irreps the others.
We will mostly use a module notation in
the following.

The
$q$-deformed classical irreps (type A) are labelled by their half-integer
spin $j$, such that $1 \le 2j+1 \le m$, and by another discrete parameter
$\omega = \pm 1$.
They are given \rRA\ by the basis
$\{w_{0},...,w_{2j}\}$ and, in a notation of module,
\eqn\eSpinj{
\cases {
k w_{p}=\omega q^{2j-2p} w_{p} \cr
f w_{p}=w_{p+1} & for $0\le p \le 2j-1$ \cr
f w_{2j}=0 \cr
e w_{p}=\omega [p][2j-p+1] w_{p-1} & for $1\le p \le 2j$ \cr
e w_{0}=0 \cr
}}
where as usual
\eqn\eQN{[x]\equiv {q^{x}-q^{-x} \over q-q^{-1}}.}
We denote this representation by $\Spin(j,\omega)$.
On it, the central elements $e^m$, $f^m$, $k^{m}$,
and $C$ take the values $x=y=0$, $z=(\omega q^{2j})^m=\pm 1$, and
$c= \omega (q-q^{-1})^{-2} \left( q^{2j+1} + q^{-2j-1} \right) $
respectively.

Note that the representation $\Spin(j,\omega=-1)$  can be obtained
as the  tensor product of $\Spin(j,1)$
by the one-dimensional representation  $\Spin(j=0,\omega)$.

\bigskip

A type B irrep is an irreducible representation that has no finite
dimensional analogue when $q$ is equal to one.
It has dimension $m$ and is characterized by three complex parameters
$\beta$, $y$, and $\lambda$.
This representation
is given in the basis $\{v_{0},...,v_{m-1}\}$ by
\eqn\eTypeB{
\cases {
k v_{p}=\lambda q^{-2p} v_{p} \cr
f v_{p}= v_{p+1} & for $0\le p \le m-2$ \cr
f v_{m-1}=y v_0 \cr
e v_{p}=\left( [p][\mu-p+1] + y\beta  \right) v_{p-1}
      & for $1\le p \le m-1$  \cr
e v_{0}= \beta v_{m-1} \cr
}}
with the definition $q^{\mu} \equiv \lambda$.

\medskip
The central elements $e^m$, $f^m$, $k^{m}$, and $C$ take the values
\eqn\eX{x=\beta \prod_{p=1}^{m-1} \left( [p][\mu-p+1] + y\beta  \right),}
$y$, $z=\lambda ^m$, and
$c= y\beta + (q-q^{-1})^{-2} \left(q \lambda + q^{-1} \lambda^{-1} \right)$
respectively.
The numbers $x$, $y$ and $z$ actually almost characterize the
irreps, up to a discrete choice for the value of the quadratic Casimir $C$
\eCas,  which satisfies an $m^{\rm th}$-degree polynomial equation with
coefficients depending on $x$, $y$ and $z$ \rDK.

We will denote the representation given by \eTypeB\ either by
$\TBi(\beta,y,\lambda)$ or equivalently by $\TBii(x,y,z,c)$. The first
notation turns out to be of much simpler use when there is a highest- or
lowest-weight vector. The second one is directly related to the spectrum of
the centre.

The representation \eTypeB\ is actually irreducible iff  one of
the three following conditions is satisfied:
\item {a. }{ $x \neq 0 $,}
\item {b. }{ $y \neq 0 $}
\item {c. }{
$\beta=0$ and $\lambda ^2 \in \CC\backslash \{1,q^2,...,q^{2(m-2)} \}$.}

\noindent
(Note that $\TBi(0,0,\pm q^{m-1})=\Spin((m-1)/2,\pm 1)$ is actually
of type A.)

The representation \eTypeB\ will be called periodic if $xy \neq 0$.
In this case it is irreducible and has no highest-weight
and no lowest-weight vector.

A semi-periodic representation is a representation for which one only of the
parameters $x$ or $y$ vanishes.  It is then also irreducible.

\newsec {Composition of type A representations}

This section will be a brief review of the results of Pasquier and Saleur
\rPS, and of Keller \rKel.

The tensor product of two representations $\Spin(j_{1},\omega_1)$
and $\Spin(j_{2},\omega_2)$
decomposes into
irreducible representations of the same type and also, if
$2(j_{1}+j_{2})+1$ is greater than $m$, into some indecomposable spin
representations \refs{\rPS ,\rKel}.

An indecomposable  spin representation $\Ind(j,\omega)$ has dimension $2m$.
It is
characterized by a half-integer $j$ such that $1\le 2j+1 <m$
and by $\omega =\pm 1$.   On a basis
$\{w_{0},...,w_{m-1}, x_{0},...,x_{m-1}\}$ the generators of
$SL(2)_{q}$ act as follows~:
\eqn\eInd{
\cases {
k w_{p}=\omega q^{-2j-2-2p} w_{p} \cr
f w_{p}=w_{p+1} & for $0\le p \le m-2$ \cr
f w_{m-1}=0 \cr
e w_{p}=\omega  [p][-2j-p-1] w_{p-1} & for $0\le p \le m-1$ \cr
k x_{p}=\omega q^{2j-2p} x_{p} \cr
f x_{p}=x_{p+1} & for $0\le p \le m-2$ \cr
f x_{m-1}=0 \cr
e x_{p}=f^{p+m-2j-2} w_{0}+ \omega [p][2j-p+1] x_{p-1}
    & for $0\le p \le m-1$\cr }}
(In particular, $e x_{0}=w_{m-2j-2}$ and $e x_{2j+1}=w_{m-1}$,
and $e^{m}$, $f^{m}$ are $0$ on such a module.)

This indecomposable representation contains the sub-representation
$\Spin(j,\omega)$. It is a deformation of the sum of the classical
$\Spin(j)$ and $\Spin(m/2-j-1)$ representations.

The fusion rules are
\eqn\eFusA {
\eqalign{
\Spin(j_1,\omega_1)\otimes \Spin(j_2,\omega_2)=&
\left( \bigoplus_{j=|j_1-j_2|} ^{\min(j_1+j_2,m-j_1-j_2-2)}
\Spin(j,\omega_1\omega_2) \right) \cr
&\bigoplus
\left( \bigoplus_{j=m-j_1-j_2-1}^{(m-1)/2} \Ind(j,\omega_1\omega_2)\right)
\;,\cr }}
where the sums are limited to integer values of $j$ if $j_ 1+j_2$ is
integer, and to half-(odd)-integer values if $j_ 1+j_2$ is
half-(odd)-integer. In conformal field theories, the fusion rules \eFusA\ are
truncated to the first parenthesis, keeping only the representations of
$q$-dimension different from 0 in the result.

The fusion rules of type A representations close with
\eqn\eFusAii{\eqalign{
&\Spin(j_1,\omega_1)\otimes\Ind(j_2,\omega_2)=
\bigoplus_{{\rm some\;} j,\omega} \Ind(j,\omega)  \cr
&\Ind(j_1,\omega_1)\otimes\Ind(j_2,\omega_2)=
\bigoplus_{{\rm some\;} j,\omega} \Ind(j,\omega)  \;.\cr
}}

\newsec {Fusion rules mixing type A and type B representations }

{\bf Proposition 1: }{\sl
The tensor product of
a semi-periodic representation with a spin $j$ representation is
completely reducible. More  precisely,
\eqn\eFusAB{
\TBi(0,y,\lambda) \otimes \Spin(j,\omega) = \bigoplus _{l=0}^{2j}
\TBi(0,(\omega(q)^{2j})^m y,q^{2(j-l)}\lambda\omega)
\;.}}
{\it Proof: }
\rDAFus\
All the weight spaces of the tensor product have the same dimension $2j+1$.
In each of them there is a singular vector, i.e. a vector in the kernel of
$\rho(e)$. Since $\rho(f)$ is injective, these vectors generate
semi-periodic sub-representations which do not mix, because the quadratic
Casimir takes different values on each of them.

{\it Remark 1:}
We stated the proposition with a highest-weight semi-periodic
representation.  A similar decomposition holds if $y=0$ and $x\neq 0$.

{\it Remark 2:}
The decomposition \eFusAB\ also holds with $y=0$, as soon as $\lambda^2$ is
not a power of $q$.

\medskip
{\bf Proposition 2: }{\sl
The tensor product of
a semi-periodic representation with
the indecomposable  spin  representation $\Ind(j,\omega)$
is completely reducible. Moreover,
\eqn\eFusABii{
\TBi(0,y,\lambda) \otimes \Ind(j,\omega) = \bigoplus _{l=0}^{m-1}
2\ \TBi(0,(\omega(q)^{2j})^m y,q^{2l}\lambda\omega)
\;.}
}
{\it Proof: }
\rDAFus\
We can find $2m$ highest weight vectors in the tensor product, or use the
previous proposition and the coassociativity of  $\Delta$.

\medskip
{\bf Proposition 3: }{\sl
The tensor product of a periodic representation $\TBii(x,y,z,c)$  with the
spin representation $\Spin(j,\omega)$ is reducible for generic values of the
parameters defining the periodic representation, and
\eqn\eFusABiii{
\TBii(x,y,z,c) \otimes \Spin(j,\omega) = \bigoplus _{l=0}^{2j}
\TBii(x,(\omega(q)^{2j})^m y,(\omega(q)^{j})^m z,c_l)
\;.}}
{\it Proof: }
One can first consider $j=1/2$. The  quadratic Casimir $C$ \eCas\ is
diagonalizable on the tensor product iff
$c\neq \pm 2/(q-q^{-1})^2$. (This defines
the generic case for $j=1/2$.) So we
obtain in this case the direct sum \eFusABiii, where $c_0$ and $c_1$ are the
(different) eigenvalues found for $C$ on the tensor product.
In the non-generic case (i.e. if $c$ takes one of the
two values $\pm 2/(q-q^{-1})^2$) the result is a (periodic) indecomposable
representation of dimension $2m$ with a quite simple structure (two copies
of \eTypeB\ with the same parameters, plus a branching from one to the
other). We then go to $j>1/2$ using the coassociativity of $\Delta$ or by a
direct analogous proof.

\newsec{Fusion of type B representations}

Consider two irreps of type B: $\rho_{1}=\TBii(x_1,y_1,z_1,c_1)$  and
$\rho_{2}=\TBii(x_2,y_2,z_2,c_2)$.

Then the central elements $e^m$, $f^m$, $k^{m}$ are scalar on
the tensor product  $\rho=(\rho_{1}\otimes \rho_{2})\circ \Delta$ and  take
the values
\eqn\exyz{
\eqalign{x&=x_1+z_1 x_2 ,       \cr
         y&=y_1 z_2^{-1}+y_2,   \cr
         z&=z_1 z_2.            \cr }}
They are also
scalar on  $\rho'=(\rho_{1}\otimes \rho_{2})\circ \Delta'$  and  take
the values  $(x'=x_2+z_2 x_1,y'=y_2 z_1^{-1}+y_1,z'=z_1 z_2)$.

We see that $\rho$ and $\rho'$ can be equivalent only if their parameters
belong to the same algebraic curve \rDJMM
\eqn\eCurve{
{x_1\over 1-z_1}={x_2\over 1-z_2}
,\qquad
{y_1\over 1-z_1^{-1}}={y_2\over 1-z_2^{-1}}}
and that in this case $x=x'$, $y=y'$, $z=z'$ also satisfy these relations.

Until the end of this section we consider the composition of $\rho_1$ and
$\rho_2$ with $\Delta$, without imposing the condition \eCurve.

Each weight space of
$\TBii(x_1,y_1,z_1,c_1)\ \otimes \ \TBii(x_2,y_2,z_2,c_2)$
has dimension $m$. The weights are all the $m^{\rm th}$ roots of $z=z_1z_2$.

The rank of $\Delta(e)$ restricted to a weight space is either
$m$ or $m-1$.
It does not depend on the weight. It is equal to
$m$ if $x\neq 0$ and to $m-1$
if  $x=0$.
\medskip
{\bf Proposition 4: }{\sl
For generic values of the parameters $(x_1,y_1,z_1,c_1)$ and
$(x_2,y_2,z_2,c_2)$, the tensor product is reducible and
\eqn\eFusBBi{
\TBii(x_1,y_1,z_1,c_1)\otimes \TBii(x_2,y_2,z_2,c_2)  =
\bigoplus _{l=0}^{m-1}
\TBii(x,y,z,c_l)
\;,}
where $x$, $y$ and $z$ are given by \exyz.}

{\it Proof: }
Consider the characteristic polynomial of the quadratic Casimir $C$ \eCas\ on
one of the weight spaces of the tensor product. The parameters $x_1$, $y_1$,
$x_2$ and $y_2$ always enter in the coefficient through the products
$x_1 y_1$ and $x_2 y_2$, except in the constant term where a non-trivial
linear combination of the products  $x_1 y_2$ and $x_2 y_1$ appears. So this
polynomial has $m$ distinct roots for generic values of the parameters.
These roots are then all the allowed values for $c$ with a given $(x,y,z)$.
Since the characteristic polynomial of $C$ is continuous in the parameters,
it is then proportional, for all the values of the parameters of the
representations, to the polynomial of ref. \rDK, the roots of which are the
possible values of $c$ for given $(x,y,z)$. Our non-generic case happens
when this polynomial has not only simple zeroes.

In the generic case,  the
eigenvectors of $C$  generate the $m$ periodic representations
$\TBii(x,y,z,c_l)$.

\medskip
{\it Remark: } in ref. \rDJMM,
the underlying quantum Lie algebra is the affine
$\widehat{SL}(N)_q$. Analogous tensor products are in this case
irreducible, in contrast with the present results. Remember that in our case
the dimension of irreps is bounded by $m$.
\medskip
{\bf Proposition 5: }{\sl
Consider values of the parameters $(x_1,y_1,z_1,c_1)$ and
$(x_2,y_2,z_2,c_2)$ such that on the tensor product $xy=0$ and $z^2\neq 1$.
Let us choose $x=0$. Then the tensor product is reducible and
\eqn\eFusBBii{
\TBii(x_1,y_1,z_1,c_1)\otimes \TBii(x_2,y_2,z_2,c_2)  =
\bigoplus _{l=0}^{m-1}
\TBii(0,y,z,c_l)
\;.}}
{\it Proof: }
Since $x=0$,   $\Delta(e)$ has rank $m-1$ on each weight space. So there is
one highest-weight vector in each weight space. Since $z^2\neq 1$, each
of them generates an $m$-dimensional representation with no singular vector.
The values of $C$ are distinct on these representations.

{\it Remarks: }
\item {1) } { Proposition 5 includes the case of the composition of
semi-periodic representations, except when $z_1z_2=\pm 1$.
This last case is more subtle. We will discuss it after the second remark.}
\item {2) } { Two tensor products of type B irreps
giving the same $(x,y,z)$ are generically equivalent: according to
proposition 4 they have the same decomposition. This is also true when the
parameters satisfy the assumptions of proposition 5. However, this is not
true for {\it all} the values of the parameters, as we will see.}
\medskip
Consider now the tensor product
$\TBii(x_1,y_1,z_1,c_1)\otimes \TBii(x_2,y_2,z_2,c_2)$
leading to $x=y=0$, $z=\pm 1$.
Note that, according to \exyz, there is still some freedom
for the choice of the parameters (on a three-dimensional manifold).
We claim
that this tensor product is equivalent, for generic values of the remaining
parameters, to the tensor product
\eqn\eFusBBiii{
\TBi(\beta_1=0,y_1=0,\lambda_1)\otimes \TBi(\beta_2=0,y_2=0,\lambda_2 )
,\qquad {\rm with} \quad (\lambda_1 \lambda_2)^m=z,}
which is a sum of indecomposable representations
$\Ind(j,\omega)$
(plus some $\Spin((m-1)/2,\omega)$).

However, there are values of the remaining parameters for which the
decomposition is not equivalent to \eFusBBiii.
For
values of the parameters lying on a submanifold (of the three-dimensional
manifold leading to  $x=y=0$, $z=\pm 1$), the invariant subspace which
generically
leads to a sub-representation equivalent to $\Ind(j,\omega)$ can now be split
into the terms of the sum
$\TBi(\beta\neq 0,y=0,\lambda=q^{2j})
\oplus \TBi(\beta\neq 0,y=0,q^{-2j-2})$.
These
representations are indecomposable and they never appear in the fusion rules
of type A irreps. They are not periodic in the sense that they correspond
to  $x=y=0$, but they share with periodic representations the fact that
$e^p$
and $f^{m-p}$ can have non-vanishing matrix elements between the same
vectors, in the basis of \eTypeB\ which diagonalizes $k$.
They contain as irreducible sub-representation the
$\Spin(m/2-j-1,-\omega)$ and $\Spin(j,\omega)$ irreps,
respectively. So the signal of this non-generic event in the already
non-generic case  $x=y=0$, $z=\pm 1$ is the appearance of
$\Spin(m/2-j-1,-\omega)$
as an irreducible sub-representation.
(Note that when $m$ is odd, $m/2-j-1$
is non-integer when $j$ is integer, and vice versa).
As an example when $m$ is even, the splitting
of the part $\Ind(j=m/2-1,\omega)$ occurs when
$c_1\ (=c_2)=0$ and $\Spin(0,-\omega)$ appears in the spectrum.

\medskip
We end this section with a remark on the regular representation of
$SL(2)_q$.  It is defined on the vector space
$\cU_q(SL(2))$ itself with the further relations $e^m=f^m=0$,
$k^m=1$, and has dimension $m^3$. It is equivalent to
$\oplus_{p=0}^m \TBi(0,0,\lambda)\otimes \TBi(0,0,\lambda^{-1}q^{2p})$.

\newsec{$\cR$-matrices}
When $q$ is a root of unity, there is no universal $\cR$-matrix
intertwining  $\Delta$ and $\Delta'$ at the level of the algebra.
When the representations $(\rho_1 \otimes \rho_2) \circ \Delta$ and
$(\rho_1 \otimes \rho_2) \circ \Delta'$ are equivalent, there exist
$\cR(1,2)$ such that
\eqn\eR{
\forall X \in SL(2)_q  \qquad
\cR  (1,2)(\rho_1 \otimes \rho_2) \circ \Delta (X)
=  (\rho_1 \otimes \rho_2) \circ \Delta ' (X) \cR (1,2)\;.}

The truncation of the formal universal $\cR$-matrix
\eqn\eRu {\cR_u= q^{-\demi h \otimes h}
\sum_{n=0}^{m-1} q^n {(1-q^2)^n \over [n]! } q^{-n(n-1)/2}
(k^{-1} e)^n \otimes  (k f)^n
}
(where $k\equiv q^h$) provides intertwiners for $\Delta$ and
$\Delta'$ when evalutated on tensor product of type A representations.
This is also true with the truncation of the inverse of the permuted
universal $\cR$-matrix
\eqn\eRui {\tilde \cR_u= q^{\demi h \otimes h}
\sum_{n=0}^{m-1} {(q-q^{-1})^n \over [n]! } q^{n(n-1)/2}
f^n \otimes  e^n
\;.}
These intertwiners satisfy Yang--Baxter equations altogether.

Tensor products with $\Delta$ and
$\Delta'$ of type B representations are not always equivalent
\refs{\rDJMM, \rGRS \rGS {--} \rRuiz}.
When the parameters of the representations
lie on the same algebraic curve \eCurve, intertwiners for these tensor
products have been found in refs. \refs{\rDJMM, \rGRS \rGS {--} \rRuiz},
in relation with the Boltzmann weight of some statistical
model. They also satisfy the Yang--Baxter equation.

We now recall results of refs. \refs{\rDAFus, \rDAslnfusion} on
$\cR$-matrices for tensor products involving both types (A and B) of
representations.  Let us call $\cR^+(1,2)$ (resp. $\cR^-(1,2)$) the
evaluation of  $\cR_u$ (resp. $\tilde \cR_u$) on the tensor product of
representations 1 and 2.

{\bf Proposition 6:} {\sl
Let $\TBii(x,y,z,c)$ and  $\TBii(x',y',z',c')$ be two representations for
which there exists an intertwiner $\cR (x,{x'} )$ ($x$ and $x'$ will
refer in the following to the whole sets of parameters $(x,y,z,c)$ and
$(x',y',z',c')$). Let   $\Spin(J,\omega)$ be a type A irrep.
We denote by $V_x$, $V_{x'}$ and $V_J$ the vectors spaces on which these
representations act.
Then the
following  Yang--Baxter equations are satisfied,
\item{a) } {
On $V_x \otimes V_{x'}  \otimes V_J$,
\eqn\eYBEi {\cR _{12}(x,{x'} )\cR^+  _{13}(x,J)\cR^+  _{23}({x'} ,J)
=\cR^+  _{23}({x'} ,J)\cR^+  _{13}(x,J)\cR _{12}(x,{x'} )\;.}
}
\item{b) } {
On $V_x \otimes V_J \otimes V_{x'}$,
\eqn\eYBEii {\cR^+  _{12}(x,J )\cR _{13}(x,x')\cR^-  _{23}(J ,x')
=\cR^-  _{23}(J ,x')\cR _{13}(x,x')\cR^+  _{12}(x,J )\;.}
}
\item{c) } {
On $V_{J}  \otimes V_x \otimes V_{x'}$,
\eqn\eYBEiii {\cR^-  _{12}(J,x )\cR^-  _{13}(J,x')\cR _{23}(x ,x')
=\cR _{23}(x ,x')\cR^-  _{13}(J,x')\cR^-  _{12}(J,x)\;.
}
}
\item{d) } { One can replace in a), b) and c) above one or both of the type B
representations $\TBii(x,y,z,c)$ and  $\TBii(x',y',z',c')$ by type A irreps,
changing $\cR (x,{x'} )$ to the
corresponding $\cR^+ $ (or also $\cR^- $), and eqs.
\eYBEi, \eYBEii, \eYBEiii\ are still
valid. Furthermore, all the type A irreps can also be replaced by
indecomposable representations occurring in the fusion rules of type A irreps.
Finally,  $\cR^+ $ and $\cR^- $ can be exchanged globally in each equation.}

{\bf However},
\item{e) } { the Yang--Baxter equation }
\eqn\eYBEiv {\cR^+  _{12}(x,J)\cR _{13}(x,{x'} )\cR^+  _{23}(J,{x'} )
=\cR^+  _{23}(J,{x'} )\cR _{13}(x,{x'} )\cR^+  _{12}(x,J)}
\item {}{ {\bf cannot} be satisfied on $V_x \otimes V_J \otimes V_{x'} $
for generic $x$ and $x'$.} }
\medskip

The pairs
$\left(\cR^+(x,J),\cR^-(J,x)\right)$ and
$\left(\cR^-(x,J),\cR^+(J,x)\right)$ are actually the only
solutions for intertwiners satisfying  Yang--Baxter equations altogether and
with $\cR(x,x')$ when periodic representations are involved.

In ref. \rDAslnfusion, a quantum chain  is presented as an
example of a new physical model involving both type A and type B
representations.

\medskip
{\bf Acknowledgements: } I thank the organizers of this beautiful conference
and the members of Trakya \"Universitesi for their hospitality.
I thank Sacha Turbiner for frequent discussions at Karaa\v{g}a\c{c}.
I also
thank \v{C}edomir Crnkovi\v{c} and
Vladimir Rittenberg for stimulating discussions at CERN.

\listrefs
\bye